# Vicinity effect of massive bodies on YBa$_2$Cu$_3$O$_{6+\delta}$ being saturated with air components

A. V. Fetisov

*Institute of Metallurgy, Ural Branch of the Russian Academy of Sciences, 101 Amundsen Str., 620016 Ekaterinburg, Russia*

We study the influence of the close-lying massive bodies on the YBCO material while it is being saturated with air components. It has been found the effect that can be interpreted as attraction/repulsion of the bodies to/from the samples of hydrated YBCO. This force interaction has been confirmed by a "direct" dynamometric method for the Earth as a maximum massive body to be used in our experiments.

As has been shown in our recent study [1], the reaction of the high-temperature superconductor YBa$_2$Cu$_3$O$_{6+\delta}$ (YBCO) with air components:

$$\text{YBa}_2\text{Cu}_3\text{O}_{6+\delta} + x\text{H}_2\text{O} + y\text{O}_2 = \text{YBa}_2\text{Cu}_3\text{O}_{6+\delta+2y}(\text{H}_2\text{O})_x, \qquad (1)$$

it turns out, can proceed in a regime of "super-high reactivity", after having prepared YBCO in a special gas atmosphere with $p_{\text{H}_2\text{O}} = 110\pm10$ Pa (processed material named "γ-YBCO"). This regime is characterized by a ~$10^2$-fold accelerated reaction (1) compared to observed for "common" YBCO. γ-YBCO has manifested no less unusual properties during further study [2]. It turns out, as γ-YBCO is saturated with air components, an increasing attraction force arises between its particles, due to which the initially dispersed material becomes a monolithic body. It has also been recorded an accelerating influence of a γ-YBCO sample on the neighboring one in the course of reaction (1), etc. The totality of effects observed has allowed to assume the existence of a certain field (let us call it conventionally as "γ-field") generated by γ-YBCO. By means of this field the oxide acts on H$_2$O and O$_2$ molecules, leading to an increase in their diffusion rate toward the γ-YBCO-sample core and a decrease of the energy barriers of their adsorption, as well as on surrounding bodies. The nature of that field, however, was not determined.

In [2], among other things, an interesting effect has been found: when the height of the γ-YBCO sample ($h$) decreases below a certain threshold value ($h_{thr} \approx r$, where $r$ is the sample's radius), the "super-high reactivity" regime of reaction (1) is replaced with the ordinary one, which is not active. The existence of $h_{thr}$ has been explained by the tendency of γ-field force lines to occupy one of the competing directions in space, choosing a state with the minimum free energy. If the result of this choice is the force lines directed toward the gas phase region, reaction (1) would be able to proceed under uncharacteristic conditions (at RT and 30% humidity). Otherwise, if the γ-field force lines are directed toward the walls of the sample container, reaction (1) cannot be realized under the specified conditions.

The aim of the paper presented below is to check the hypothesis we have made in the work [2]. And the effect of switching between two reactivity regimes of reaction (1) is used as the main research method. The change in the experimental value of $h_{thr}$ is considered here as a measure of the effect of surrounding bodies on the γ-YBCO sample by means of γ-field, because the energy of such effect is clearly to change the above state with the minimum free energy. The main advantage of the method is that all its parameters to be measured are determined by simple weighing, which does not require ultra-high precision[1]. At the same time, the sensitivity of that method far exceeds the direct measurement of the force of interaction of bodies by dynamometric method, as will be shown below.

The γ-YBCO material for this study was prepared by the method described in [1, 2]. The resulting powder was poured into quartz containers with a diameter of 7.5 mm and a length of 13 mm. Containers were covered with three layers of aluminum foil of 40 μm thickness; several small holes were made in the foil to allow air to come into the container. Aluminum foil provided shielding of the sam-ples from external electric fields. In order for the environmental conditions not to affect the course of reaction (1), each container with γ-YBCO was placed into a sealed glass bottle (so-called "weighing bottle") with moisture holding material in it (providing a humidity of 30%). To study the effect of various bodies on reaction (1), the bottles were usually placed on them as shown in Fig. 1. The method of monitoring the development of reaction (1) was periodic weighing the shielded container with sample on an analytical balance Shimadzu AUW-120D (Japan) with a precision of $1 \cdot 10^{-2}$ mg. The parameter $h_{thr}$ was calculated with the empirical formula: $h_{thr}[\text{mm}] = 7.3 \cdot 10^{-3} (m_s[\text{mg}] + 69)$, where $m_s$ is the initial mass of the sample.

---

[1] The parameter $h_{thr}$ was calculated from the γ-YBCO sample mass, for which a precision of 1 mg was quite enough. The presence of the "super-high reactivity" regime of reaction (1) could be reliably established when the change of the sample mass was determined with a precision of 0.1 mg (for the case $r$ = 3.75 mm).

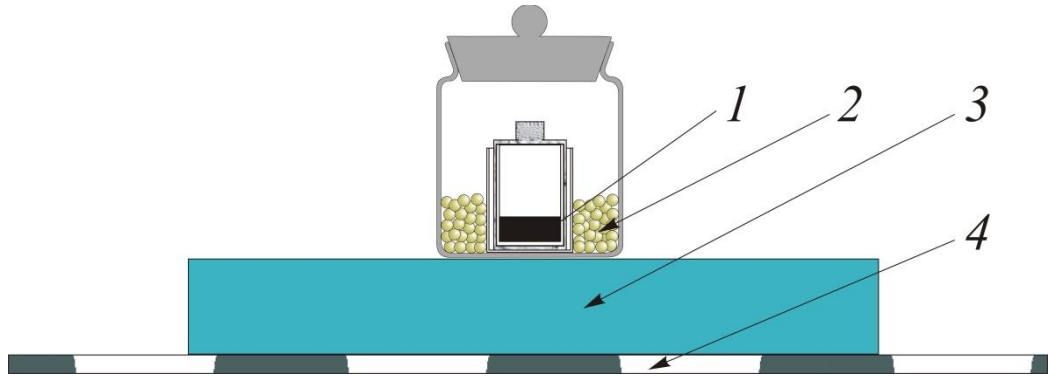

*1* – γ-YBCO sample; *2* – moisture holding material; *3* – body affecting the sample; *4* – supporting cardboard stand.

**Fig. 1.** Scheme of the experiment on the influence of various bodies on the development of reaction (1).

Curve *1* in Fig. 2 shows the result of a "blank" experiment in which the massive body affecting the sample is absent (the bottles with samples are placed just on the cardboard, see Fig. 1). It is not difficult to see that the regime of reaction (1) is sharply changed when $h$ reaches the value $h_{thr} \approx 2.46 \pm 0.13$ mm. This result differs from that obtained in [2], however, experimental conditions here and in the work [2] have been also different. In [2], the saturation of γ-YBCO with air components has been carried out in a stainless steel chamber consisting of a massive base and thin walls. As will be shown below, such a chamber design promotes reaching higher values of $h_{thr}$. The value of $h_{thr}$ corresponding to the "blank" exper-iment, in turn, has likely been formed as a result of the interaction between γ-YBCO and both quartz container walls and the Earth (see below).

The effect on samples of a permanent magnet, whose force lines enter the samples from below and a field gradient in the sample region is about 1 T/m, is shown in Fig. 2 as curves *2* and *3*. This effect leads to an increase of $h_{thr}$ up to 3.20±0.14 mm while the south pole of the magnet is directed to the sample and to 2.68±0.13 mm in the case of the north pole.

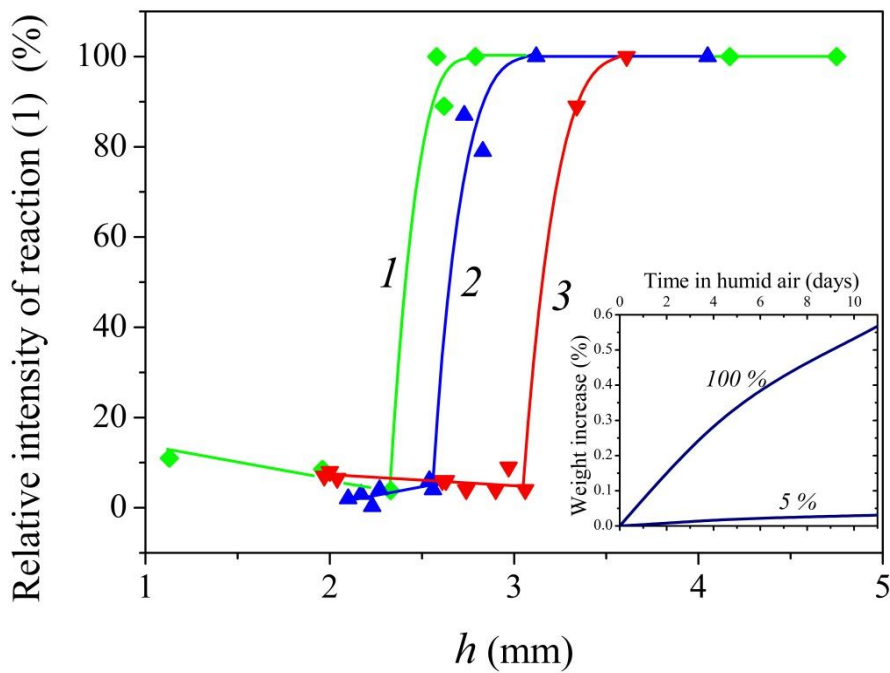

**Fig. 2.** The dependence of the critical parameter $h_{thr}$, which defines the change in the rate-regime of reaction (1), on the action of the N-pole (curve *2*) and the S-pole (curve *3*) of a permanent magnet on the γ-YBCO samples. Curve *1* corresponds to the case when there are no any external effects exerting on the samples. Inset: Examples of the reaction intensity of 100 and 5%.

Curves *2–4* in Fig. 3 reflect the influence on the γ-YBCO samples of stainless steel bodies. In the case of the single stainless steel disc located below the sample this effect leads to an increase of $h_{thr}$ to 2.9±0.2 mm. The two-sided influence on the sample of two discs – located above and below of the sample – leads to a twofold increase in $h_{thr}$, compared with the case of the single body. On the other hand, the lateral effect of the stainless steel ring on γ-YBCO leads to a significant decrease in $h_{thr}$. Curve *1* in the following Fig. 4 shows the influence on the γ-YBCO samples of the quartz rods.

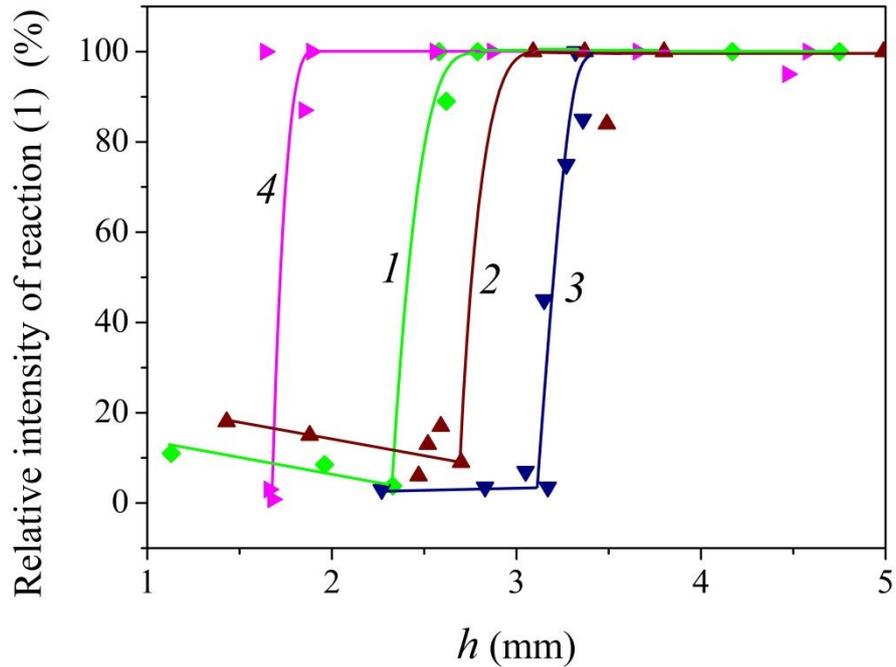

**Fig. 3.** The dependence of the critical parameter $h_{thr}$, which defines the change in the speed regime of reaction (1), on the action of stainless steel bodies on the γ-YBCO samples: disc located beneath the sample (curves *2*), two discs located beneath and above the sample (curve *3*) and ring surrounding the sample (curve *4*). Curve *1* corresponds to the "blank" experiment.

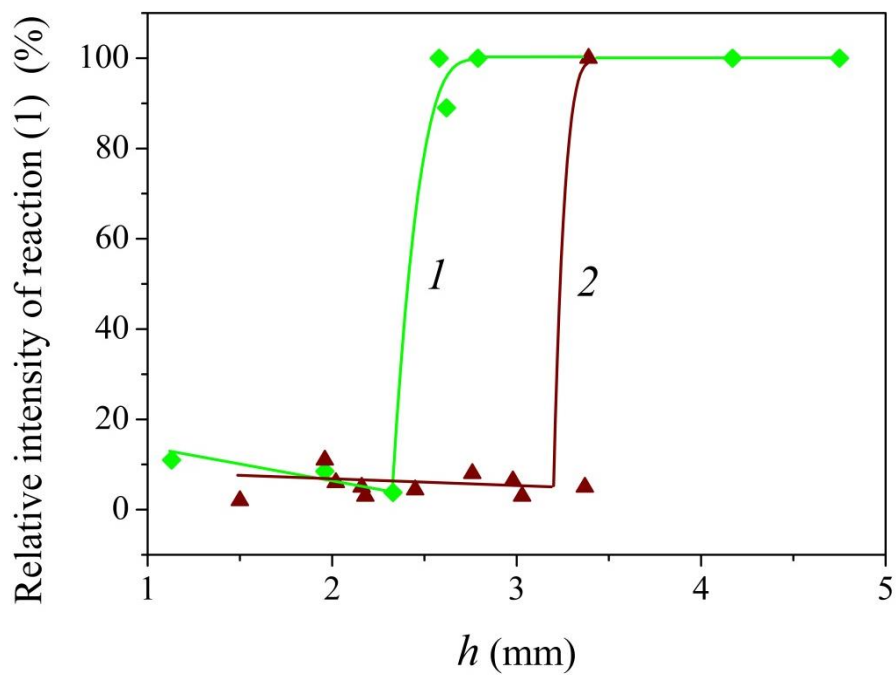

**Fig. 4.** The dependence of the critical parameter $h_{thr}$ on the action of quartz rods bundle on the γ-YBCO samples (curve *2*). Curve *1* corresponds to the "blank" experiment.

Lastly, the complete set of the results on the influence of various materials on $h_{thr}$ is summarized in Table 1.

**Table 1.** The vicinity effect of massive bodies on γ-YBCO

| Massive body affecting the γ-YBCO sample | Relative arrangement of the specific massive body (bodies) and the sample | $h_{thr}$, mm |
|---|---|---|
| None | – | 2.46±0.13 |
| A disc of non-magnetic stainless steel (70 mm diameter and 11 mm thickness) | plate is beneath sample | 2.90±0.20 |
| Two discs of non-magnetic stainless steel (70 mm diameter and 11 mm thickness) | one plate is beneath sample and the other is above it | 3.22±0.05 |
| A ring of non-magnetic stainless steel (20 mm inner diameter, 35 mm outer diameter, and 5 mm thickness) | sample is at the center of ring | 1.77±0.09 |
| A bundle of quartz rods (height of bundle is 18 mm, area is 150×500 mm$^2$) | plate is beneath sample | 3.35±0.02 |
| The S-pole of magnet (dimensions of magnet: thickness is 5 mm, area is 12×25 mm$^2$) | – " – | 3.20±0.14 |
| The N-pole of magnet (dimensions: 5×12×25 mm$^3$) | – " – | 2.68±0.13 |
| A second γ-YBCO sample in a quartz container (sample mass is about 200 mg) | bottom of the influencing sample container touches a lateral wall of the influenced sample container | > 10 |

The experimental results represented in Figs. 2–4 and in Table 1 will be discussed here from the viewpoint of the power field generated by YBCO [2] (substantially by γ-YBCO). These results imply that massive bodies as well as the S-pole of magnet placed near γ-YBCO are able to interact with the power field of γ-YBCO, giving an additional advantage to one of possible directions of its power lines. At this, if the power lines of γ-YBCO are directed to the gas phase, reaction (1) proceeds intensively, otherwise it doesn't. However, such an interpretation of the experiments may seem frivolous. And we are aware that additional very weighty arguments are required to confirm the existence of the interaction between γ-YBCO and surrounding bodies.

To solve this issue, measurements of the interaction force between the γ-YBCO sample and the S-pole of magnet by the "direct" dynamometric method, the scheme of which is depicted in Fig. 5(a), were performed. In order to record that interaction at least at $1 \cdot 10^{-7}$ n (it is the accuracy of the used method), we had to use an electromagnet producing a field gradient of about 5 T/m in the sample region, instead of the permanent magnet used in the experiments of Fig. 2. Such a replacement would lead to a 25-fold increase in the strength of interaction between the sample and the magnet. As shown in the experiment, Fig. 5(b), the interaction turns out to be a periodic function of time and it is attractive in the beginning till about 12 days, and repulsive then. In turn, under the same conditions of the magnetic experiment sketched in Fig. 5(a), the interaction between γ-YBCO and the N-pole of magnet has been not recorded. It is in agreement with the data of Fig. 2 and Table 1. However, this result is very strange.

We also tried to perform the direct measurements of the interaction force between γ-YBCO and the Earth (the latter was used as the most representative member of the family of massive bodies). To do this, we carried out periodic weighing of a sealed glass vessel with the γ-YBCO sample inside; the vessel was similar to the one depicted in Fig. 1. In the absence of mass transfer with the environment, the change in the vessel weight with time (in the course of reaction (1)) should indicate the character and magnitude of the interaction of γ-YBCO with the Earth. Fig. 5(b) shows this interaction as a function of time. It is not difficult to see that the dependence obtained here has the same periodic character as the one obtained for the S-pole of magnet.

It is worth to note that in the force of experimental restrictions the first points (non-null) on dependences in Fig. 5 have been measured only by the second day after the experiment started. And, as one can see, these points lay in the "attractive force" area. However, while acting on the samples from below, such a force should cause a decrease in $h_{thr}$. It is because this force would promote for the γ-field lines to be directed to the gas phase region and, consequently, to facilitate reaction (1). In fact, however, one observes an increase in the value of $h_{thr}$ (see Figs. 2–4). So, we assume that, immediately after beginning the experiment, that force will, after all, be a repulsive force, as shown in Fig. 5 by dotted lines.

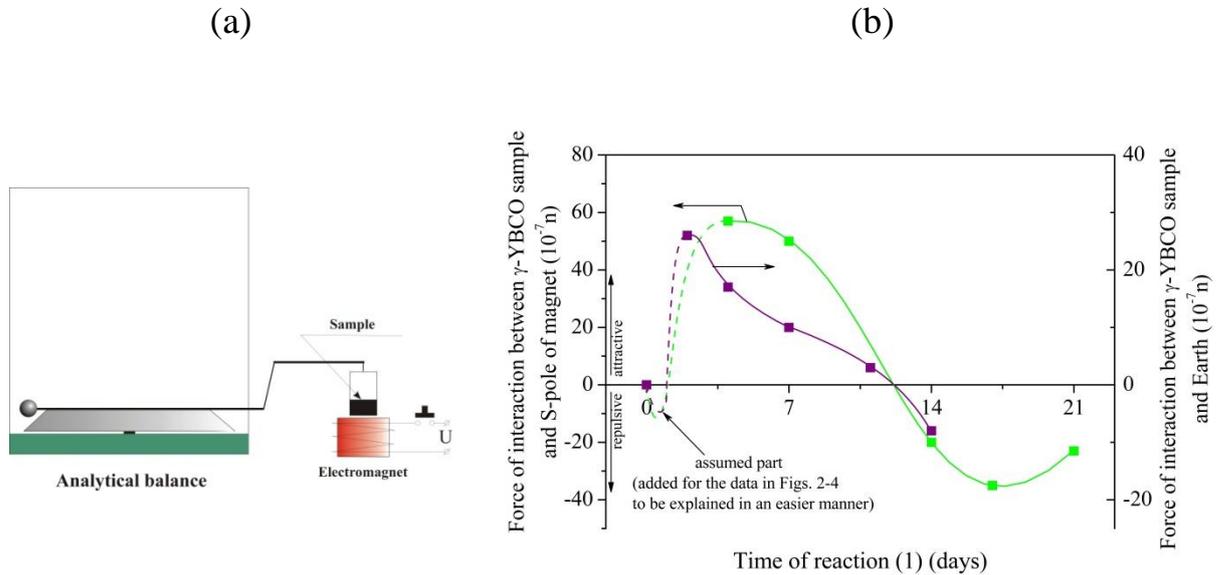

**Fig. 5.** Scheme of the experimental setup for the direct measurements of the interaction force between the S-pole of magnet and the γ-YBCO sample (a). The results of direct measurements of the interaction force between the γ-YBCO samples and the surrounding bodies: the S-pole of magnet and the Earth (b).

Further, as seen in Table 1, there is a similar influence on γ-YBCO of different bodies, except for the second γ-YBCO sample. At the latter case, in opposite to the others, the lateral influence leads to facile blocking of reaction (1) (see Table) and the influence from below leads to promoting this reaction [2]. Meanwhile, we have observed that the powdered sample of γ-YBCO gradually transform to a monolithic body along all the reaction time. All that points out that only attractive force exerts between particles as well as samples of γ-YBCO.

Based on the results of the present work, now we can try to give an alternative explanation of the effect of "gravitational force shielding by bulk YBa$_2$Cu$_3$O$_{7-x}$ superconductor" described in [3] and known from literature as the "Podkletnov effect". In the experiments of Ref. 3, a loss in weight of a small sample of SiO$_2$ suspended above a massive superconducting disc of YBCO was occurred. In our experiments, it has been observed the effect that is, in principle, equivalent to the "Podkletnov effect": the bundle of quartz rods (SiO$_2$) has placed beneath the γ-YBCO sample leads to increased $h_{thr}$ that can be interpreted in terms of the existence of the repulsive interaction between γ-YBCO and SiO$_2$. In this connection, we assume that the superconducting disc used in [3] could accidentally be treated just like as described in [1, 2] and got properties inherent to the γ-YBCO material.

Thus, in the present work, we test the influence of the close-lying massive bodies on the new γ-YBCO material. The result obtained is interpreted in terms of the existence of a certain force field acting between them.